# Excitons in epitaxially grown $WS_2$ on Graphene: a nanometer-resolved EELS and DFT study

Max Bergmann[1,*], Jürgen Belz[1], Oliver Maßmeyer[1], Badrosadat Ojaghi Dogahe[1], Robin Günkel[1], Johannes Glowatzki[1], Andreas Beyer[1], Ivan Solovev[2], Jens-Christian Drawer[2], Martin Esmann[2], Sergej Pasko[3], Simonas Krotkus[3], Michael Heuken[3], Stefan Wippermann[1] and Kerstin Volz[1,*]

[1]Material Sciences Center and Department of Physics, Philipps-Universität Marburg, Germany

[2] Institute of Physics, Carl von Ossietzky University of Oldenburg, Germany

[3] AIXTRON SE, Herzogenrath, Germany

* Corresponding authors:
max.bergmann@physik.uni-marburg.de
kerstin.volz@physik.uni-marburg.de

**Abstract**

**In this study, we investigate excitonic properties of epitaxially grown $WS_2$, which is of particular interest for various applications due to its potential for upscaling to wafer-sized structures. Understanding the effect of the dielectric environment due to changing layer numbers and multi-material heterostructures on the optical properties is crucial for tailoring device properties. Monochromated electron energy loss spectroscopy in a scanning transmission electron microscope is employed to characterize the excitonic spectrum of $WS_2$ on graphene grown by metal organic chemical vapor deposition. This technique provides the required spatial resolution at the nanometer scale in combination with high quality spectra. To complement the experimental results, theoretical investigations using density functional theory and applying the Bethe-Salpeter equations are conducted. We find that by transitioning from mono- to bi- to multilayers of $WS_2$ the spectra show redshifts for both, the K-valley excitons at about 2.0 and 2.4 eV as well as excitonic features of higher energies. The latter features originate from so-called band nesting of transitions between the Γ- and K-point. In summary, this study provides valuable insights into the excitonic properties of $WS_2$ in different layer configurations and environments, which are realistically needed for future device fabrication and property tuning. Finally, we can show that nanometer-scale electron spectroscopy supported by careful theoretical modelling can successfully link atomic structure and optical properties, such as exciton shifts, in non-idealized complex material systems like multilayer 2D heterostructures.**

**Keywords: 2D, TMDs, excitons, 2D heterostructures, stacking, MOCVD, STEM, EELS, $WS_2$, graphene, bilayer**

## Introduction

In the field of materials science, two-dimensional (2D) transition metal dichalcogenides (TMDs) have emerged as a captivating area of research with remarkable potential, particularly in the domain of valleytronics[1–7]. TMDs form a diverse class of 2D materials, with semiconducting properties making them ideal for various applications like battery electrodes[8], solar cells[9,10] and field-effect tunneling transistors[11,12]. TMDs also feature an indirect-to-direct band gap transition in the monolayer (ML) limit, which leads to a vastly enhanced quantum efficiency and significant increase in photoluminescence (PL)[13–15], thus illustrating their potential for optoelectronic applications. The optoelectronic applications of these TMDs are fundamentally linked to the behavior of excitons in the proximity of the K valley within the first Brillouin zone. One unique feature of TMDs' electronic properties is their tunability through layer number variation, what is of particular interest. Additionally, TMDs such as $WS_2$ possess spin–orbit interactions that couple the spin and valley degrees of freedom. Understanding the impact of local structural variations on the electronic properties is crucial to tailor possible device applications. Especially, in the example of field-effect tunneling transistors, a heterostructure of $WS_2$ and graphene, that is also investigated in this work, is shown to be a promising candidate.

The role of excitons, quasiparticles of electron-hole pairs bound by Coulomb force, is essential in low-dimensional systems like ML TMDs[16]. These systems are characterized by enhanced Coulomb interactions due to spatial confinement and reduced Coulomb screening[17,18]. While the existence of valley excitons in ML TMDs has been theoretically predicted[19], also various techniques have been explored, including inelastic x-ray scattering[20], electron energy loss spectroscopy (EELS)[21] and neutron scattering[22] to address the experimental observation of excitons. However, these methods have their limitations and none is entirely suitable for probing the intrinsic exciton properties of TMDs.

In this context, EELS, primarily conducted in monochromated scanning transmission electron microscopes (STEMs), is a promising avenue to investigate the elusive exciton properties in ML TMDs. STEM-EELS not only offers the required spatial resolution to visualize atomic structures and potentially local defects and correlate the excitonic signal, but also yields the probability to supply a momentum[23] in contrast to spectroscopic techniques reliant on phonons, enabling indirect transitions. This capability arises from the markedly greater momentum carried by electrons compared to photons.

In this study, we employ STEM-EELS to investigate the energy shifts of K-valley excitons and contrast them with energy shifts of excitons residing at higher energies. This investigation pertains to the impact of local structural variations that arise as a consequence of transitioning from a ML to a bi- or multilayer (BL, MuL) configuration. While other studies focus on moiré-twisted TMD samples[24,25], we investigate an AA-stacked structure that originates from metal organic chemical vapor deposition (MOCVD) grown $WS_2$ on a graphene substrate. There are several studies on the excitonic excitations originating from the K-K gap[25–28] and the corresponding spin-orbit coupling, often referred as the A and B excitations. However, none of these studies have investigated the energy shifts of this excitations due to number of layers using EELS and also neither of them has used a technologically highly relevant, grown heterostructure. This work is yet to overcome this and compares the magnitude of the energy shift of excitons residing at the K-point with that of higher energy excitonic effects residing at other points in the first Brillouin zone. First principle calculations, namely Density Functional Theory (DFT) with the Bethe-Salpeter equations (BSE) on top[29–32], were used to interpret the spectra and explain the energy shifts.

**Methods**

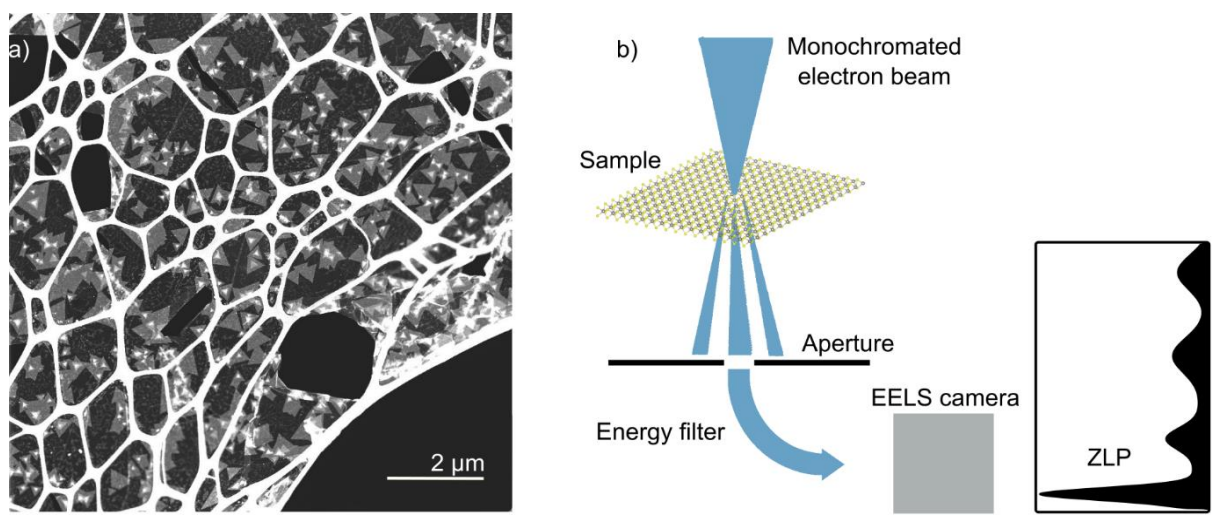

**Figure 1: a)** Overview of the freestanding $WS_2$ on graphene sample on the lacey-carbon grid. The latter is the honeycomb-like structure that serves as a support for the thin 2D material. The triangles are the $WS_2$ flakes laying on top of a graphene layer, that gives slightly more contrast in comparison to the vacuum in the bottom right of the picture. **b)** Schematic of a STEM-EELS setup.

The $WS_2$-graphene heterostructure under study is synthesized using metal-organic chemical vapor deposition (MOCVD) in a 19x2 inch Close Coupled Showerhead AIXTRON MOCVD reactor. The process is conducted on c-plane sapphire substrates, which possess a 0.2° off-cut towards the m-plane. The fabrication of the heterostructure started with the deposition of graphene on the sapphire substrate. This is achieved using methane as a precursor at a temperature of 1400°C, incorporating a surface pre-treatment step involving hydrogen, as detailed in reference [33]. Following this, $WS_2$ nucleation is carried out on the graphene-coated sapphire template. This process utilizes tungsten hexacarbonyl ($W(CO)_6$) and ditertiarybutylsulfide (DTBS) as precursors, with a growth temperature set at 700°C and a growth duration of 1800 seconds, as described in [34,35].

Subsequently, an etchant-free transfer technique is employed to relocate the specimen onto a lacey carbon STEM grid, as visually depicted in Fig. 1a. This results in a free-standing sample spanning the holes between the carbon support. In the initial stage of this transfer process, polymethyl-methacrylate (PMMA) is spin-coated onto the sapphire substrate. Subsequently, the specimen is immersed in water heated to a temperature of 80°C. This immersion results in the intercalation of water between the thin film and the substrate, facilitating the film's detachment from the substrate. The 2D film is then collected onto the TEM grid and the PMMA is subsequently dissolved using dichloromethane. The process is schematically depicted in Fig. S1a in the supplementary part.

High angle annular dark-field (HAADF) images are taken using a double aberration-corrected JEOL 2200FS STEM operated at 80 kV with a semi-convergence angle of 21 mrad. In HAADF mode high-angle Rutherford-like scattered electrons are detected. Notably, the scattering cross section of these electrons is proportional to the atomic number (Z) of the sample material, following an empirical relationship of between $Z^{1.5}$ and $Z^2$ depending on the detector angle. This relation forecasts higher contrast for the W-Atoms in comparison to the S-Atoms as well as an increased brightness for increasing number of layers. The EELS-data is collected using a monochromated JEOL JEM-ARM200F NEOARM STEM operated at 60 kV applying a double-Wien filtered monochromator limiting the energy width of the probe to 35 meV. A probe semi-convergence angle of 1 mrad is chosen resulting in a spot size of 6 nm. Using an aperture in the diffraction plane a momentum transfer can be selected. We choose an aperture of the size of 300 µm indicated in Fig. S1b by the red circle. A Gatan GIF Continuum HR, an energy filter followed by an EELS camera, is then used to record the spectrum. The setup is schematically depicted in Fig. 1b. The EELS data is evaluated using the Hyperspy python package[36].

In order to understand the origin of the experimentally observed signals, we perform DFT calculations as implemented in the GPAW-package[37–39] and the Atomic Simulation Environment (ASE)[40]. The electron-ion interaction is described via projector augmented wave pseudopotentials[41] and the Perdew-Burke-Ernzerhof (PBE)[42] exchange-correlation functional. To correct for the underestimation of the electronic gap by DFT-PBE, a single-shot PBE0[43,44] calculation is performed on top of the DFT-PBE electronic structure (cf. Fig. S2 for a comparison to GW results). The spin-orbit coupling is then calculated non-self-consistently. To account for the weak van-der-Waals forces between layers during the relaxation process for the BL, we use the DFT-D3 dispersion correction[45,46] in conjunction with Becke-Johnson damping[47]. The computed EELS spectra include excitonic effects at the level of many-body perturbation theory via solving the BSE. In the BSE calculations, the Brillouin zone integration is performed using a Monkhorst-Pack mesh corresponding to 30x30x1 k-points in the primitive $WS_2$ unit cell. Moreover, our BSE calculations use a 2D truncation scheme[48,49] in order to exclude Coulomb-based interactions between periodic images of the supercells in the direction normal to the $WS_2$ surface. The BSE is solved using the DFT-PBE0 electronic structure as an input. We converge the EELS spectra with respect to the number of occupied and empty states entering the BSE calculations up to an energy of 3.25 eV, achieved by including 6 topmost valence bands and 6 lowest conduction bands of the $WS_2$ system.

**Results and discussion**

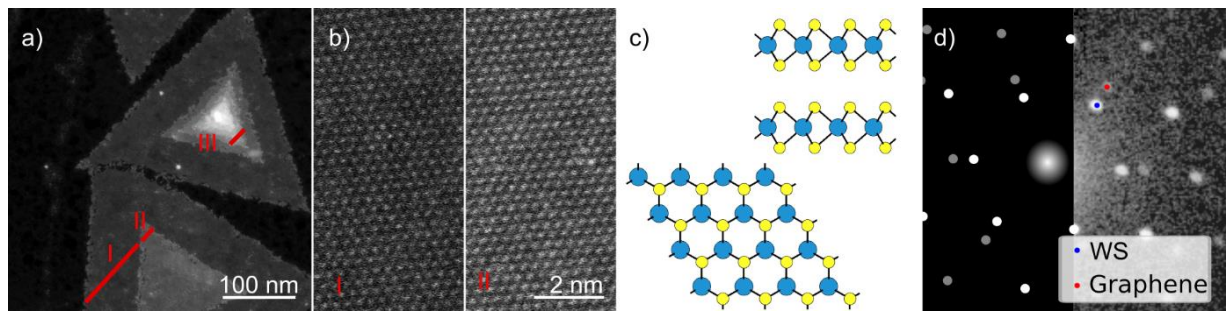

**Figure 2: a)** Two $WS_2$ flakes on graphene with regions of mono- bi- and multilayered material, which can be distinguished by their brightness. Red lines indicate the EELS scan lines labeled as I, II and III for the different number of layers: ML, BL and MuL. **b)** Regions I and II depicted in high resolution. **c)** The upper schematics illustrates the AA-stacking configuration observed in region II, while the lower schematics presents a top view of AA-stacked $WS_2$, highlighting that in projection no difference, except for the increased intensity, is evident compared to ML $WS_2$. **d)** Nanobeam diffraction pattern from region I. The left part of the image is a schematic continuation of the diffraction pattern to guide the eye. The diffraction pattern unambiguously reveals both $WS_2$ and graphene diffractions, providing conclusive evidence of the graphene substrate as well as the heteroepitaxial alignment of the $WS_2$ to the graphene.

We first discuss the EELS data acquisition and the structure of the corresponding scanned sample region, followed by an evaluation of the EEL spectra in comparison with simulation results.

EELS data acquisition involves conducting systematic line scans across two distinct $WS_2$ flakes containing mono-, bi-, and multilayered structures. The specific scanning trajectory is depicted by the red line in Fig. 2a. Fig. 2a itself is generated through HAADF imaging, showing increased brightness with increasing number of layers allowing to count the number of WS$_2$ layers. Fig. 2b shows a cutout of the high-resolution images of ML region I and the BL region II. The heavy W-atoms have brighter contrast than the light S-atoms. By comparing the intensities and atomic configurations in Fig. 2b, one can determine the stacking order within the BL region to be AA-stacking (schematically depicted in Fig. 2c), thereby facilitating the later alignment of ab-initio calculations with this specific stacking configuration. Next to as well as underneath the WS$_2$ flakes we find the diffractions of graphene, as shown in Fig. 2d. This confirms the heteroepitaxial alignment of the WS$_2$ on the graphene achieved during MOCVD growth.

The EELS spectrum, as presented in Fig. 3a, is acquired by positioning an aperture around the forward-scattered electron spot, like indicated in Fig. S1b by the red circle, allowing for momentum transfers covering a distance of roughly 75 % of the distance between Γ and K. The spectra are each acquired at the denoted positions of ML, BL and MuL (I, II and III in Fig. 2a), smoothed by Savitzky-Golay filtering[50] and averaged over all scan points on the corresponding regions. Moreover, the elimination of the zero-loss peak's (ZLP) residual tail, originating from additional scattering phenomena, e.g. radiation losses, is executed by employing a double exponential decay model. Additionally, a dedicated scan of the region exclusively containing the graphene substrate is executed. Subsequently, this substrate-specific data is subtracted from each of the spectra denoted as I, II, and III, effectively eliminating the contribution from graphene and ensuring a more focused analysis of the WS$_2$ spectrum. However, the electronic influence of the graphene on the WS$_2$ of course cannot be subtracted easily, and hence is part of further discussions in this work.

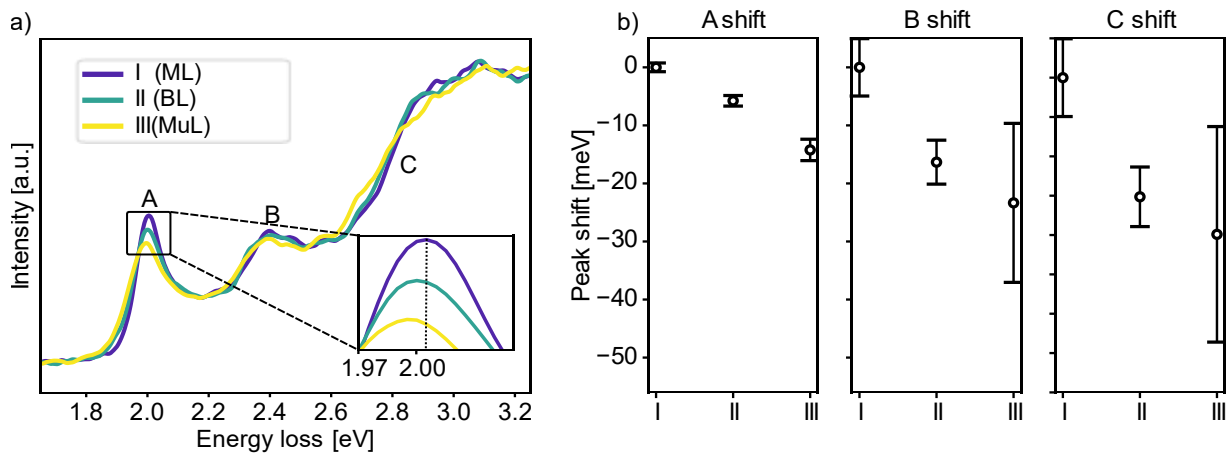

**Figure 3: a)** EEL spectrum of the three distinct regions labeled as I, II and III in Fig. 2a. The most prominent features are labeled as A, B and C. The inset shows the redshift of the A peak around 2eV. **b)** Peak shifts relative to the peak position of ML region I for the three features A, B and C shown in Fig. 3a. For the error bars multiple random selections out of the corresponding region are fitted and their standard deviation is taken as error bar.

The dataset manifests two prominent excitonic peaks positioned at 2.0 eV and 2.4 eV, which we assign as A and B, in accordance with prior work measuring optical reflectance[26]. Additionally, an edge 'C' is detected, spanning the energy range of 2.7 eV to 2.9 eV. The inset within Fig. 3a highlights the A exciton peak position at different regions I, II, and III, revealing a notable redshift with increasing layer number. The A peak, which corresponds to an optically bright transition is also successfully detected using micro-photoluminescence spectroscopy, as shown in Fig. S3 in the supplementary part. It is observed throughout the entire sample, underlining that our observations are a general property of the heterostructure. To quantitatively assess the spectral changes in the EELS data depending on layer number, an analytical fit to the entire spectrum shown is performed. This fit encompasses five Gaussian distributions in conjunction with a *tanh* function applied to the data within each respective region, thereby facilitating the precise determination of the energy positions of the distinct peaks and the edge C. Further elaboration on this fitting procedure can be found in the supplementary part Fig. S4. Fig. 3b then displays the shifts of the A and B peaks, as well as of the edge C. A conspicuous redshift is evident for all features with increasing layer number. While a redshift in the order of a few millielectronvolts (meV) characterizes the lower energy features, the redshift of a few tens of meV is more pronounced for the higher energy features, like C. Additionally, a different slope of the C-feature is detected in region III (Fig. 3a) in comparison to spectra I and II.

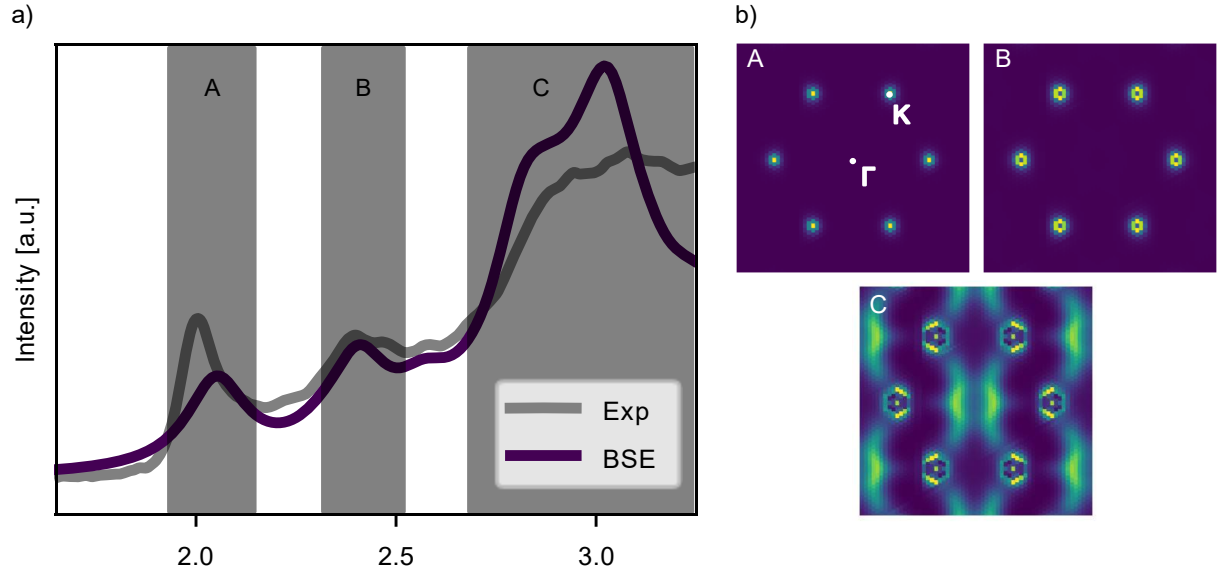

**Figure 4: a)** Results of the BSE simulations for ML $WS_2$ in comparison to the experimental result. **b)** Transition matrix elements in the ML BSE simulation for the three different peaks indicated by the gray boxes in Fig. 4a.

To achieve a comprehensive understanding of the experimental results, we now turn to DFT-BSE calculations. We model the regions I and II shown in Fig. 2b, respectively, first as freestanding $WS_2$ MLs and freestanding AA-stacked BLs without the graphene substrate. The computed EELS spectrum of the ML (cf. Fig. 4a.) shows two prominent excitonic features labeled A and B. Except for a small blueshift of the simulation of 50 meV in comparison to the experiment, these features are consistent with the experimental observations both in terms of the absolute energy, as well as the peak distance of 0.35 eV due to the spin-orbit coupling[51] at the K-point of the Brillouin zone. The spectral features at higher energies beyond 2.5 eV (C peak in the simulated spectra), also match the experimental observations.

As the presence of the graphene substrate can be important for device applications[11,12], it is imperative to understand the influence of the graphene substrate on the electronic structure of the $WS_2$ layers. We hypothesize, that its presence contributes to the experimentally observed EELS spectra, either indirectly via dielectric effects[52] or directly by modifying the electronic structure, and may well contribute to the observed 50 meV shift between the experiment and DFT calculations. The presence of graphene is understood to induce notable band shifts[52] due to decreasing quantum confinement, leading to a reduction of the band gap. Moreover, the additional dielectric screening provided by the graphene substrate decreases the binding strength between excitonic electron-hole pairs in $WS_2$, leading to more diffuse excitonic features and shifting them to slightly higher energies. The complex interplay between these contrasting effects affects both the shape and the position of the excitonic peaks. In total a redshift of the A peak has been shown theoretically and experimentally by Waldecker *et al.*[52].

We now investigate the dependence of the EEL spectra on the number of $WS_2$ layers. As we transition into the BL or MuL regions, the $WS_2$ layers, shown for instance within region II and III in Fig. 2a, also act as mutually interacting dielectric environments. Our present EELS measurements show a redshift of the A exciton with increasing number of $WS_2$ layers, consistent with reflectance spectroscopy measurements[26,53]. In contrast to purely optical techniques, however, our EELS measurements have a lateral resolution on a nanometer length scale and thus clearly correlate the energy shift of the A exciton to an AA-stacked BL. We also attribute this redshift to the decreasing quantum confinement and hence decreasing band gap with increasing number of layers. Consistent with recent quasiparticle calculations in GW approximation and solving the Wannier equations[52,54], these band shifts slightly overcompensate the decreasing exciton-binding energy, that would otherwise shift the A exciton to higher energies. To clarify the origin of the features in the simulation and hence the experiment, we show the transition matrix elements in Fig. 4b, specifically the eigenvectors belonging to the BSE Hamiltonian, corresponding to the three prominent peaks labelled as A, B and C in Fig. 4a. A and B originate mainly from excitations at the K-point, indicating their association with the spin-orbit split K-valley excitons already known from literature[51]. In contrast, the origins of the C excitations appear more multifaceted. Points between Γ and K within the Brillouin zone contribute significantly, which has been attributed to band nesting effects reported in previous studies[55]. As these resonances do not only originate from the K-point, an examination of the band structures of $WS_2$ ML and BL material, as illustrated in Fig. 5a and 5b respectively, can provide valuable insights. Those transitions illustrated in Fig. 4b, that contribute most prominently to the A, B and C peaks, are marked by arrows in the band structures shown in Fig. 5a. The interaction between the layers leads to a lifting of the degeneracy of bands away from the effective monolayer band structure, leading to splits in the band structure. At the K-point, we observe a minimal split of ~50 meV of the bands and a reduction in the band gap due to the BL configuration of 26 meV. In contrast, at points between Γ and K a more pronounced split is clearly visible. These findings explain both the measured redshift of the A resonance, as well as the more pronounced redshift of the C feature, since these transitions stem from points between Γ and K (cf. Fig. 4b). Hence, the reduced exciton binding energy is overcompensated by the shrinking electronic gap. As more layers enter the system, more bands with lifted degeneracy contribute to the excitations, particularly at points between Γ and K, leading to a more diverse spectrum. This may provide an explanation for the altered slope of the C feature in region III.

The practical implementation of the aperture in the experiment (see Fig. S1b) allows momentum transfers. Thus, it becomes imperative to consider higher q-contributions in the simulation, as shown in the supplementary Fig. S5. These contributions are rather small and decay rapidly with increasing q for contributions in the vicinity of the A and B peak. Energy contributions above 2.7 eV are more pronounced and the cumulative effect over the Brillouin zone could potentially modify the spectrum, providing an additional explanation for the discrepancies between simulation and experiment in the vicinity of 3.25 eV. Of course, future in-depth studies focusing on the contributions at finite momentum transfers should be carried out, possibly revealing information about dark excitons. These excitons are likely to be directly observable with EELS due to the momentum transferred by the electrons.

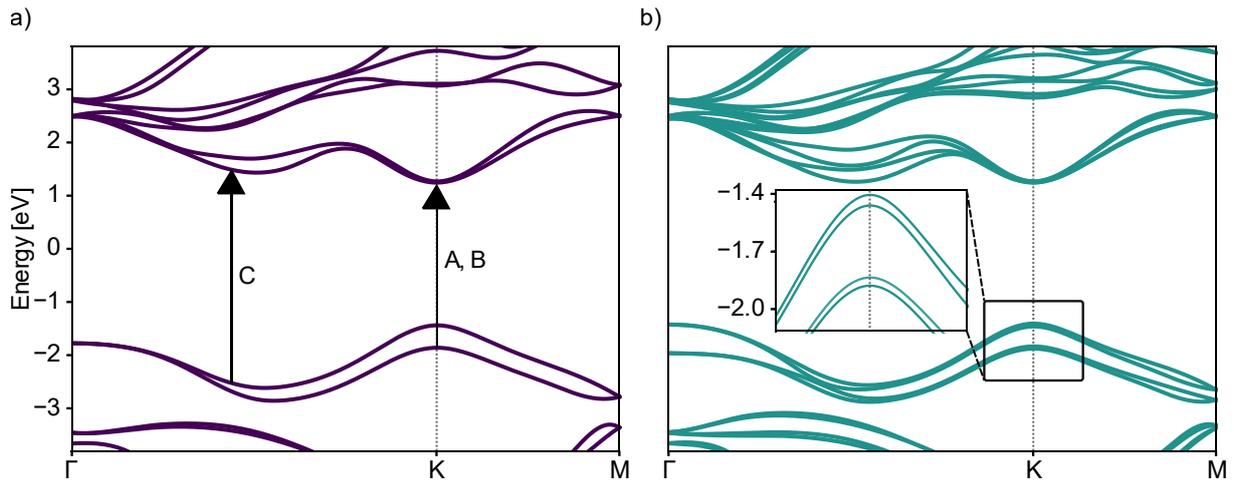

**Figure 5: a)** DFT band structure of ML WS$_2$. The arrows show the transitions in momentum space, where the prominent features A, B and C appear. **b)** DFT band structure of BL WS$_2$. A splitting of the bands is visible at the K-point, leading to a reduction of the band gap.

**Conclusion**

Via joint nanometer-scale EEL spectroscopy and *ab initio* simulations, we investigate the energy shift characteristics of the WS$_2$ K-valley excitons, which exhibit a redshift during the transition from ML to AA-stacked BL and MuL configurations. The determination of the AA stacking configuration was obtained through the acquisition of a high-resolution STEM image within the specified region. Particularly, the redshift becomes more pronounced when examining features with energy levels exceeding 2.5 eV. Employing *ab initio* simulations, we assigned the measured EEL spectral features to specific transitions in the WS$_2$ monolayer and bilayer band structures. Notably, the C feature, that stems from an intermediate point between Γ and K, is considerably more susceptible to variations in the layer configuration, thereby explaining the observed dissimilarities between the shifts. Our investigations correlate excitonic features in WS$_2$ over a wide energy range with structural variations, thereby providing novel pathways toward layered materials with tailored properties for device applications. As our sample is grown on graphene, our findings are of particular interest for the field-effect tunneling transistor[11,12].

**Acknowledgement**

We acknowledge financial support from the German Research Foundation (DFG) in the framework of the collaborative research center SFB 1083 "Structure and Dynamics of Internal Interfaces" (project number: 223848855). We thank Dr. Philipp Wachsmuth and Dr. Masaki Mukai (JEOL GmbH) for their help operating the monochromated NeoARM microscope used for this study.

References

1. Wang, G. *et al.* Colloquium : Excitons in atomically thin transition metal dichalcogenides. *Rev. Mod. Phys.* **90,** 21001; 10.1103/RevModPhys.90.021001 (2018).

2. Raja, A. *et al.* Coulomb engineering of the bandgap and excitons in two-dimensional materials. *Nature communications* **8,** 15251; 10.1038/ncomms15251 (2017).

3. Zeng, H., Dai, J., Yao, W., Di Xiao & Cui, X. Valley polarization in MoS2 monolayers by optical pumping. *Nature nanotechnology* **7,** 490–493; 10.1038/nnano.2012.95 (2012).

4. Cao, T. *et al.* Valley-selective circular dichroism of monolayer molybdenum disulphide. *Nature communications* **3,** 887; 10.1038/ncomms1882 (2012).

5. Zheng, S. *et al.* Coupling and Interlayer Exciton in Twist-Stacked WS 2 Bilayers. *Advanced Optical Materials* **3,** 1600–1605; 10.1002/adom.201500301 (2015).

6. Mak, K. F., He, K., Shan, J. & Heinz, T. F. Control of valley polarization in monolayer MoS2 by optical helicity. *Nature nanotechnology* **7,** 494–498; 10.1038/nnano.2012.96 (2012).

7. Rycerz, A., Tworzydło, J. & Beenakker, C. W. J. Valley filter and valley valve in graphene. *Nature Phys* **3,** 172–175; 10.1038/nphys547 (2007).

8. David, L., Bhandavat, R. & Singh, G. MoS2/graphene composite paper for sodium-ion battery electrodes. *ACS nano* **8,** 1759–1770; 10.1021/nn406156b (2014).

9. Fortin, E. & Sears, W. M. Photovoltaic effect and optical absorption in MoS2. *Journal of Physics and Chemistry of Solids* **43,** 881–884; 10.1016/0022-3697(82)90037-3 (1982).

10. H. Tributsch. Layer-Type Transition Metal Dichalcogenides — a New Class of Electrodes for Electrochemical Solar Cells (1977).

11. Novoselov, K. S., Mishchenko, A., Carvalho, A. & Castro Neto, A. H. 2D materials and van der Waals heterostructures. *Science (New York, N.Y.)* **353,** aac9439; 10.1126/science.aac9439 (2016).

12. Zhang, S. *et al.* Two-dimensional heterostructures and their device applications: progress, challenges and opportunities—review. *J. Phys. D: Appl. Phys.* **54,** 433001; 10.1088/1361-6463/ac16a4 (2021).

13. Eda, G. *et al.* Photoluminescence from chemically exfoliated MoS2. *Nano letters* **11,** 5111–5116; 10.1021/nl201874w (2011).

14. Splendiani, A. *et al.* Emerging photoluminescence in monolayer MoS2. *Nano letters* **10,** 1271–1275; 10.1021/nl903868w (2010).

15. Mak, K. F., Lee, C., Hone, J., Shan, J. & Heinz, T. F. Atomically thin MoS$_2$: a new direct-gap semiconductor. *Physical review letters* **105,** 136805; 10.1103/PhysRevLett.105.136805 (2010).

16. Chernikov, A. *et al.* Exciton binding energy and nonhydrogenic Rydberg series in monolayer WS(2). *Physical review letters* **113,** 76802; 10.1103/PhysRevLett.113.076802 (2014).

17. Olsen, T., Latini, S., Rasmussen, F. & Thygesen, K. S. Simple Screened Hydrogen Model of Excitons in Two-Dimensional Materials. *Physical review letters* **116,** 56401; 10.1103/PhysRevLett.116.056401 (2016).

18. Komsa, H.-P. & Krasheninnikov, A. V. Effects of confinement and environment on the electronic structure and exciton binding energy of MoS 2 from first principles. *Phys. Rev. B* **86,** 241201; 10.1103/PhysRevB.86.241201 (2012).

19. Coehoorn, R., Haas, C. & Groot, R. A. de. Electronic structure of MoSe2, MoS2, and WSe2. II. The nature of the optical band gaps. *Physical review. B, Condensed matter* **35,** 6203–6206; 10.1103/PhysRevB.35.6203 (1987).

20. Tornatzky, H., Gillen, R., Uchiyama, H. & Maultzsch, J. Phonon dispersion in MoS2. *Phys. Rev. B* **99,** 144309; 10.1103/PhysRevB.99.144309 (2019).

21. Reidy, K. *et al.* Direct Visualization of Subnanometer Variations in the Excitonic Spectra of 2D/3D Semiconductor/Metal Heterostructures. *Nano letters* **23,** 1068–1076; 10.1021/acs.nanolett.2c04749 (2023).

22. Wakabayashi, N., Smith, H. G. & Nicklow, R. M. Lattice dynamics of hexagonal Mo S2 studied by neutron scattering. *Phys. Rev. B* **12,** 659–663; 10.1103/PhysRevB.12.659 (1975).

23. Schuster, R., Wan, Y., Knupfer, M. & Büchner, B. Nongeneric dispersion of excitons in the bulk of WSe2. *Phys. Rev. B* **94,** 85201; 10.1103/PhysRevB.94.085201 (2016).

24. Susarla, S. *et al.* Mapping Modified Electronic Levels in the Moiré Patterns in MoS2/WSe2 Using Low-Loss EELS. *Nano letters* **21,** 4071–4077; 10.1021/acs.nanolett.1c00984 (2021).

25. Woo, S. Y. *et al.* Excitonic absorption signatures of twisted bilayer WSe2 by electron energy-loss spectroscopy. *Phys. Rev. B* **107,** 155429; 10.1103/PhysRevB.107.155429 (2023).

26. Li, Y. *et al.* Measurement of the optical dielectric function of monolayer transition-metal dichalcogenides: MoS2 , MoSe2 , WS2 , and WSe2. *Phys. Rev. B* **90,** 205422; 10.1103/PhysRevB.90.205422 (2014).

27. Nerl, H. C. *et al.* Probing the local nature of excitons and plasmons in few-layer MoS2. *npj 2D Mater Appl* **1,** 1–9; 10.1038/s41699-017-0003-9 (2017).

28. Hong, J., Senga, R., Pichler, T. & Suenaga, K. Probing Exciton Dispersions of Freestanding Monolayer WSe_{2} by Momentum-Resolved Electron Energy-

Loss Spectroscopy. *Physical review letters* **124,** 87401; 10.1103/PhysRevLett.124.087401 (2020).

29. Rohlfing, M. & Louie, S. G. Electron-Hole Excitations in Semiconductors and Insulators. *Phys. Rev. Lett.* **81,** 2312–2315; 10.1103/PhysRevLett.81.2312 (1998).

30. Hybertsen, M. S. & Louie, S. G. Electron correlation in semiconductors and insulators: Band gaps and quasiparticle energies. *Physical review. B, Condensed matter* **34,** 5390–5413; 10.1103/PhysRevB.34.5390 (1986).

31. Albrecht, S., Reining, L., Del Sole, R. & Onida, G. Ab Initio Calculation of Excitonic Effects in the Optical Spectra of Semiconductors. *Phys. Rev. Lett.* **80,** 4510–4513; 10.1103/PhysRevLett.80.4510 (1998).

32. Benedict, L. X., Shirley, E. L. & Bohn, R. B. Optical Absorption of Insulators and the Electron-Hole Interaction: An Ab Initio Calculation. *Phys. Rev. Lett.* **80,** 4514–4517; 10.1103/PhysRevLett.80.4514 (1998).

33. Mishra, N. *et al.* Wafer-Scale Synthesis of Graphene on Sapphire: Toward Fab-Compatible Graphene. *Small (Weinheim an der Bergstrasse, Germany)* **15,** e1904906; 10.1002/smll.201904906 (2019).

34. Djordje Dosenovic, Samuel Dechamps, Celine Vergnaud, Sergej Pasko & Hanako Okuno. *Mapping domain junctions using 4D-STEM: toward controlled properties of epitaxially grown transition metal dichalcogenide monolayers* (2023).

35. Tang, H. *et al.* Nucleation and coalescence of tungsten disulfide layers grown by metalorganic chemical vapor deposition. *Journal of Crystal Growth* **608,** 127111; 10.1016/j.jcrysgro.2023.127111 (2023).

36. Francisco de la Peña *et al. hyperspy/hyperspy: Release v1.7.3* (Zenodo, 2022).

37. Enkovaara, J. *et al.* Electronic structure calculations with GPAW: a real-space implementation of the projector augmented-wave method. *Journal of physics. Condensed matter : an Institute of Physics journal* **22,** 253202; 10.1088/0953-8984/22/25/253202 (2010).

38. Mortensen, J. J., Hansen, L. B. & Jacobsen, K. W. Real-space grid implementation of the projector augmented wave method. *Phys. Rev. B* **71,** 35109; 10.1103/PhysRevB.71.035109 (2005).

39. Yan, J., Mortensen, J. J., Jacobsen, K. W. & Thygesen, K. S. Linear density response function in the projector augmented wave method: Applications to solids, surfaces, and interfaces. *Phys. Rev. B* **83,** 245122; 10.1103/PhysRevB.83.245122 (2011).

40. Hjorth Larsen, A. *et al.* The atomic simulation environment-a Python library for working with atoms. *Journal of physics. Condensed matter : an Institute of Physics journal* **29,** 273002; 10.1088/1361-648X/aa680e (2017).

41. Blöchl, P. E. Projector augmented-wave method. *Physical review. B, Condensed matter* **50,** 17953–17979; 10.1103/PhysRevB.50.17953 (1994).

42. Perdew, J. P., Burke, K. & Ernzerhof, M. Generalized Gradient Approximation Made Simple. *Physical review letters* **77,** 3865–3868; 10.1103/PhysRevLett.77.3865 (1996).

43. Adamo, C. & Barone, V. Toward reliable density functional methods without adjustable parameters: The PBE0 model. *J. Chem. Phys.* **110,** 6158–6170; 10.1063/1.478522 (1999).

44. Perdew, J. P., Ernzerhof, M. & Burke, K. Rationale for mixing exact exchange with density functional approximations. *J. Chem. Phys.* **105,** 9982–9985; 10.1063/1.472933 (1996).

45. Grimme, S., Antony, J., Ehrlich, S. & Krieg, H. A consistent and accurate ab initio parametrization of density functional dispersion correction (DFT-D) for the 94 elements H-Pu. *The Journal of chemical physics* **132,** 154104; 10.1063/1.3382344 (2010).

46. Grimme, S., Ehrlich, S. & Goerigk, L. Effect of the damping function in dispersion corrected density functional theory. *Journal of computational chemistry* **32,** 1456–1465; 10.1002/jcc.21759 (2011).

47. Smith, D. G. A., Burns, L. A., Patkowski, K. & Sherrill, C. D. Revised Damping Parameters for the D3 Dispersion Correction to Density Functional Theory. *The Journal of Physical Chemistry Letters* **7,** 2197–2203; 10.1021/acs.jpclett.6b00780 (2016).

48. Rozzi, C. A., Varsano, D., Marini, A., Gross, E. K. U. & Rubio, A. Exact Coulomb cutoff technique for supercell calculations. *Phys. Rev. B* **73,** 205119; 10.1103/PhysRevB.73.205119 (2006).

49. Hüser, F., Olsen, T. & Thygesen, K. S. How dielectric screening in two-dimensional crystals affects the convergence of excited-state calculations: Monolayer MoS 2. *Phys. Rev. B* **88,** 245309; 10.1103/PhysRevB.88.245309 (2013).

50. Savitzky, A. & Golay, M. J. E. Smoothing and Differentiation of Data by Simplified Least Squares Procedures. *Anal. Chem.* **36,** 1627–1639; 10.1021/ac60214a047 (1964).

51. Zhu, Z. Y., Cheng, Y. C. & Schwingenschlögl, U. Giant spin-orbit-induced spin splitting in two-dimensional transition-metal dichalcogenide semiconductors. *Phys. Rev. B* **84,** 153402; 10.1103/PhysRevB.84.153402 (2011).

52. Waldecker, L. *et al.* Rigid Band Shifts in Two-Dimensional Semiconductors through External Dielectric Screening. *Physical review letters* **123,** 206403; 10.1103/PhysRevLett.123.206403 (2019).

53. Zhao, W. *et al.* Evolution of electronic structure in atomically thin sheets of WS2 and WSe2. *ACS nano* **7,** 791–797; 10.1021/nn305275h (2013).

54. Rösner, M., Şaşıoğlu, E., Friedrich, C., Blügel, S. & Wehling, T. O. Wannier function approach to realistic Coulomb interactions in layered materials and heterostructures. *Phys. Rev. B* **92,** 85102; 10.1103/PhysRevB.92.085102 (2015).

55. Carvalho, A., Ribeiro, R. M. & Castro Neto, A. H. Band nesting and the optical response of two-dimensional semiconducting transition metal dichalcogenides. *Phys. Rev. B* **88,** 115205; 10.1103/PhysRevB.88.115205 (2013).

## Supplementary Information

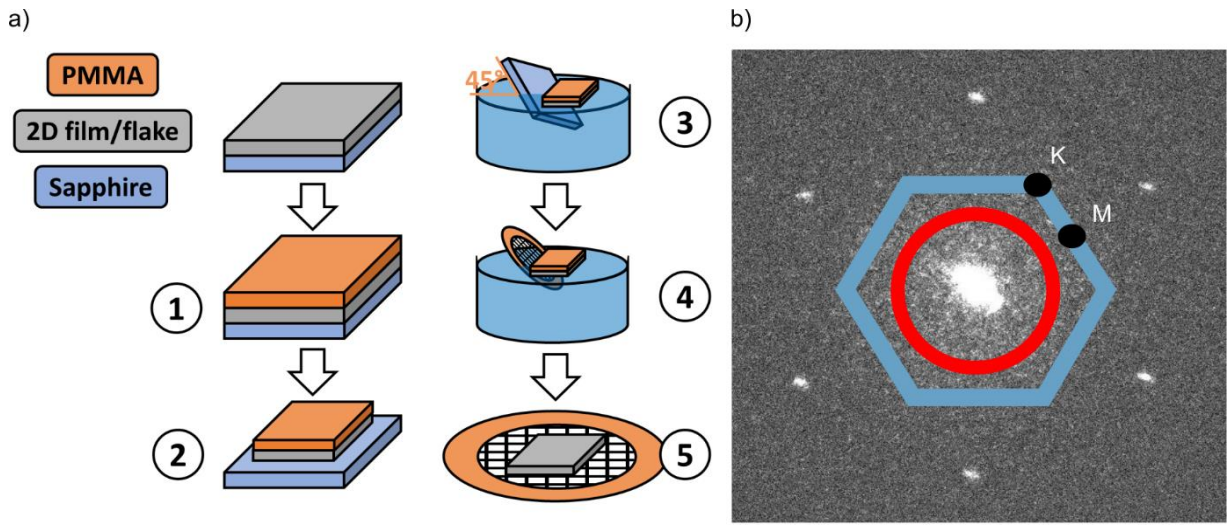

**Figure S1: a)** PMMA transfer. **b)** Image of the diffraction plane as obtained in the experiment. The red circle indicates how the aperture is set in comparison to the Brillouin zone, indicated by the blue hexagon.

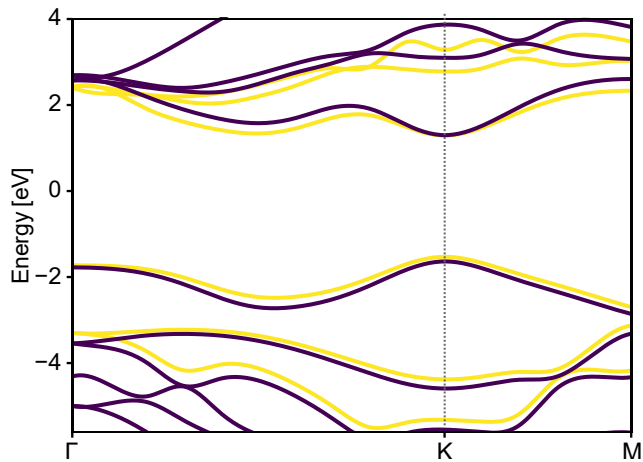

**Figure S2:** Comparison between the band structure obtained using DFT+PBE0 and the result obtained with the GW approach without spin-orbit coupling. A slight discrepancy in the band gap at K is visible. Also note the strong discrepancy at points between Γ and K, which leads to a shift of the C-feature to higher energies in the BSE spectrum when using PBE0, thus providing a better agreement with the experiment.

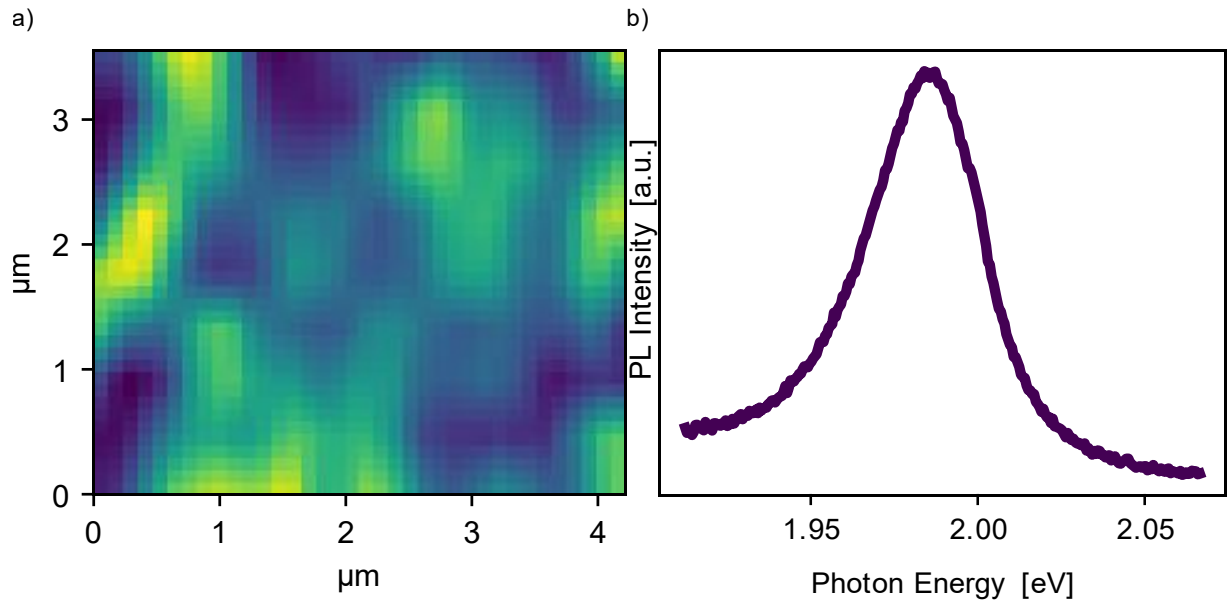

**Figure S3:** a) The PL map of a random sample region reveals different PL intensities of the A peak depending on the number of flakes in the given scan spot. b) Average PL spectrum of the map in a). One obtains a strong peak at 1.985 eV, which fits to the A peak obtained in the EELS study.

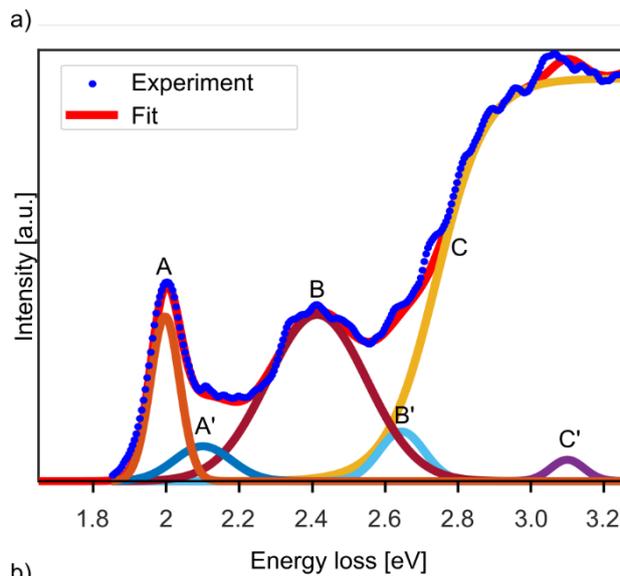

| Position on Sample | A | A' | B | B' | C | C' |
|---|---|---|---|---|---|---|
| Counts [A.U.] | | | | | | |
| I | 4,32E-05 | 1,84E-05 | 1,39E-04 | 3,07E-05 | 5,81E-04 | 6,33E-06 |
| II | 4,00E-05 | 1,60E-05 | 1,34E-04 | 4,41E-05 | 5,46E-04 | 7,17E-06 |
| III | 4,02E-05 | 1,46E-05 | 1,33E-04 | 4,50E-05 | 5,03E-04 | 7,87E-06 |
| x [eV] | | | | | | |
| I | 2,0004 | 2,1023 | 2,4196 | 2,6405 | 2,7649 | 3,1101 |
| II | 1,9946 | 2,1026 | 2,4045 | 2,6390 | 2,7486 | 3,1067 |
| III | 1,9862 | 2,1014 | 2,3997 | 2,6439 | 2,7416 | 3,1517 |
| σ [eV] | | | | | | |
| I | 0,0365 | 0,0723 | 0,1287 | 0,0773 | 7,1566 | 0,0632 |
| II | 0,0363 | 0,0714 | 0,1272 | 0,0801 | 7,9933 | 0,0526 |
| III | 0,0458 | 0,0775 | 0,1389 | 0,0873 | 6,8051 | 0,0670 |

**Figure S4:** Fitting procedure. a) Exemplaric fit of one of the EELS spectra in the monolayer region. Five gaussian peaks (A, A', B, B', C') in addition to one *tanh* function (C) have been found to fit the spectrum. b) Average parameters of the different features denoted in a) in the three different regions I, II and III as shown in Fig. 2a.

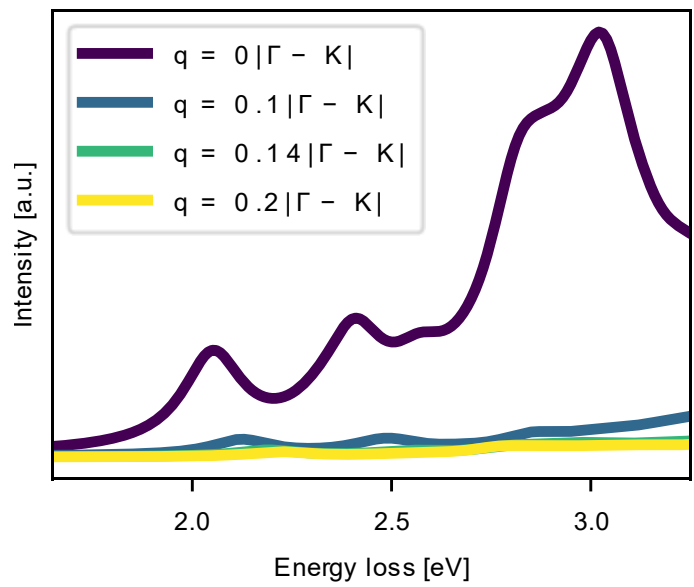

**Figure S5:** Additional contributions by simulation from finite q close to the q=0 contributions compared to the experimental spectrum of the monolayer. For increasing q, the contribution to the spectrum shrinks strongly.